# Signatures of novel magnon-phonon coupling in frustrated double perovskite square lattices

Shalini Badola, Aprajita Joshi, Akriti Singh, and Surajit Saha*

*Indian Institute of Science Education and Research Bhopal, Bhopal 462066, India*

*Correspondence: surajit@iiserb.ac.in*

**Abstract:** Low-dimensional frustrated magnetic square networks feature a variety of unconventional phases with novel emergent excitations. Often these excitations are intertwined and manifest into intriguing phenomena, an area that has remained largely unexplored in square-lattice systems, especially, double perovskites ($A_2BB'O_6$). In this study, we explore these interactions between the fundamental excitations such as phonons and magnons in square-lattice $Sr_2CuTeO_6$, $Sr_2CuWO_6$, and $Ba_2CuWO_6$ isostructural double perovskites that exhibit both short-ranged ($T_S$) as well as long-ranged Néel antiferromagnetic ($T_N$) transitions. Our Raman measurements at variable temperatures reveal an intriguing broad peak (identified as 2-magnon (2M)) surviving beyond $T_S$ for W-based compositions contrary to the Te-based system, suggesting a key role of diamagnetic B′-site cation on their magnetism. The thermal response of 2M intriguingly shows signatures of correlation with phonons and control over their anharmonicity, depicting magnon-phonon interaction. Further, a few phonons exhibit anomalies across the magnetic transitions implying the presence of spin-phonon coupling. In particular, the phonon modes at ~ 194 cm$^{-1}$ of $Sr_2CuTeO_6$ and ~ 168 cm$^{-1}$ of $Sr_2CuWO_6$, that show a strong correlation with the 2M, exhibit the strongest spin-phonon coupling suggesting their roles in mediating magnon-phonon interactions in these systems.

# I. INTRODUCTION

Frustrated square lattice (SL) cuprates have driven enormous research interest due to their close association with unconventional quantum phases and emergent spin excitations [1,2]. Frustration in SLs arrives with reduced magnetic dimensionality, enriching the magnetic phase diagram with novel states [3]. The reduced magnetic dimensionality induces pronounced quantum fluctuations which not only influence the spectrum of magnetic and lattice excitations significantly but also may promote novel spin textures (*e.g.* skyrmions), spin-liquids, superconductors, *etc* [4-7]. Investigation of spin and lattice excitations can, thus, be an indirect approach to developing a profound understanding of the properties of low-dimensional SL. Moreover, exploring SL cuprates is necessary from the standpoint of entangled quasi-particle dynamics as well as novel phases to realize applications in spintronics, and caloritronics. Although SLs have been explored theoretically in great detail, only a few model systems could be realized in practice so far to achieve profound experimental insights. One of such model systems belong to the Te/W-based class of double perovskites ($A_2BB'O_6$ : A, B-divalent cations; B' – hexavalent cation).

The B-site ordered double perovskites, $Sr_2CuTeO_6$, $Sr_2CuWO_6$ and $Ba_2CuWO_6$, host uniform SL network of $Cu^{2+}$-ions in the *ab*-plane by crystallizing into the tetragonal structure (S.G.: *I4/m*) [8-10]. The elongation of the $CuO_6$ octahedra along apical direction due to Jahn-Teller distortion of $Cu^{2+}$-ion weakens the out-of-plane (*i.e.*, *c*-axis) exchange pathway, thus, restricting the dominant magnetic interactions to the *ab*-plane. As a consequence, these systems nurture quasi-two-dimensional (2D) antiferromagnetic (AFM) character which manifests in the magnetic susceptibilities as broad magnetic transition ($T_S$). A three-dimensional (3D) ordering is further followed at lower temperatures ($T_N$) due to reduced Cu-Cu distances and increased interaction in the apical direction. More interestingly, recent studies have established that exchange interactions can be tuned to stabilize spin-liquid-like correlations in $Ba_2CuWO_6$ with chemical substitution [11]. The signatures of fractionalized spin excitations have been proposed in $Sr_2CuWO_6$ with inelastic neutron scattering measurements [12]. Therefore, such findings indicate that SLs are inherently rich in novel phases and excitations. Moreover, previous investigations on $Sr_2CuTeO_6$ and $Sr_2CuWO_6$ demonstrated that their magnetic excitations (magnons) survive beyond $2T_N$ due to their low-dimensional magnetic characters [9]. The presence of magnons beyond $T_N$ at such high temperatures have drawn immense interest in the community, the answer to which may lie with various coupled degrees of freedom. For instance, magnon-phonon coupling may govern the relaxation and decay processes of these excitations [13]. Such prospects of intertwined effects between spin and lattice excitations and different degrees of freedom (lattice, spin, charge, and orbital) have remained largely unexplored in these SLs and, thus, demand a detailed investigation.

Since inelastic light scattering is efficient in capturing the signatures of fundamental spin excitations along with their interactions with the crystal lattice [14,15], we have performed systematic temperature- and magnetic field-dependent Raman measurements on polycrystals of frustrated SLs $Sr_2CuTeO_6$, $Sr_2CuWO_6$, and $Ba_2CuWO_6$ that are synthesized using solid-state reaction. Firstly, we have addressed the presence of an intriguing broad Raman peak in all the three compositions mainly in the magnetic phase, and established it as two-magnon excitation. Our Raman measurements reveal a gradual disappearance of the two-magnon feature in both $Sr_2CuTeO_6$ and $Sr_2CuWO_6$, especially above the $T_S$. However, this feature exists beyond room temperature in $Ba_2CuWO_6$ corroborating a stronger exchange interaction ($J$) in $Ba_2CuWO_6$. Our temperature-dependent phonon analysis finds a strong correlation between the response of closely lying two-magnon and phonon excitations in $Sr_2CuTeO_6$, $Sr_2CuWO_6$, and $Ba_2CuWO_6$ across the short-range ordering temperature regime, thus implying the presence of magnon-phonon coupling.

Besides, a few phonons exhibit anomalies near the magnetic transition in all these systems suggesting the presence of spin-phonon coupling. Magnetic measurements corroborate our findings.

## II. SYNTHESIS AND METHODS

Polycrystalline samples of $A_2CuB'O_6$ (A: Ba, Sr, B: W, Te) were grown using solid-state synthesis route. The powdered precursors such as $BaCO_3$, $SrCO_3$, CuO, and $TeO_2$ were mixed in the desired proportions, ground, and treated at temperatures 900 °C and 1000 °C for 12 hrs each under ambient conditions in box furnace. The powdered samples were then palletized which is followed by a thermal treatment at 1100 °C for another 12 hrs. The phase purity and crystal structure of the samples were determined using an Empyrean PANalytical X-Ray Diffractometer which reveals the presence of a minor fraction ~ 2 % and ~ 0.7 % of the secondary phase of $AWO_4$ (A = Ba/Sr) for both $Sr_2CuWO_6$ and $Ba_2CuWO_6$, repectively. On the other hand, $Sr_2CuTeO_6$ turned out to be phase pure. Energy Dispersive X-Ray (EDAX) set up from ZEISS was employed to determine the elemental stoichiometry. Raman measurements were performed using an HR Evolution LabRAM Raman spectrometer with laser excitations of 532 nm in the backscattered configuration. The sample temperature for measurement was varied using a liquid-nitrogen-cooled Linkam THMS (Model: HFS600E-PB4) sample stage attached to a Peltier-cooled CCD detector. Additionally, a closed-cycle attoDRY 1000 cryostat was employed to perform Raman measurements down to 4 K with variable magnetic field (0 - 9T). Magnetic measurements on the synthesized samples were performed using Quantum Design SQUID-VSM set up with an applied field of 500 Oe.

## III. RESULTS AND DISCUSSION

### A. Structure and Magnetism

Rietveld refinement of the X-ray diffraction (XRD) patterns for the three compositions $Sr_2CuTeO_6$ (SCT), $Sr_2CuWO_6$ (SCW), and $Ba_2CuWO_6$ (BCW) suggest their tetragonal crystal symmetry (*I4/m*) at room temperature with lattice parameters matching with previous reports [8,9]. The XRD data are shown in the supplemental material (refer to Fig. S1) [16]. While the Sr-based compositions (SCT and SCW) have similar lattice parameters, BCW possesses a relatively larger unit cell due to the larger size of Ba-cation. EDAX measurements suggest that all the systems crystallize in the expected stoichiometry with the desired elemental proportion, as shown in the supplemental material [16]. A similar crystal symmetry of the three systems is also suggestive of the closely related magnetic and phonon properties, which are discussed below.

Figure 1 displays the magnetization data for the three compositions revealing a similar magnetic behavior. All the systems are characterized by two magnetic transitions – a Néel order ($T_N$) exhibited as a sharp anomaly in the magnetization at lower temperatures (below 30 K) and a short-ranged order ($T_S$) manifested as broad maxima at higher temperatures (near 100 K). Unlike SCW and BCW, the Néel order ($T_N$) could not be clearly deciphered for SCT from the magnetization data and, therefore, deduced using the derivative of the magnetization as presented in the inset in Fig. 1. The $T_N$ for SCW, SCT, and BCW are 26 K, 28 K, and 29 K while their $T_S$ are 100 K, 73 K, and 110 K, respectively, thus matching with previous reports (see Table SI in supplemental material) [8,9]. The presence of the AFM transition implies that finite three-dimensional long-ranged magnetic interactions emerge at low temperatures in these systems. The short-range correlations, developed by the overlapping of empty $Cu^{2+}$ $d_{x^2-y^2}$ orbitals, are highly two-dimensional in nature and accompanied by a gradual decrease in the magnetic moment with increasing temperature, particularly for BCW and SCW. On the contrary, SCT exhibits a rapid decrement in the moment above the $T_S$ indicating relatively weaker magnetic interactions. This observation is consistent with the strength of exchange interactions (*J's*) estimated from our Raman measurements (discussed later). Our Raman measurements reveal interesting findings across these transitions which are discussed in the sections below.

### B. Raman spectroscopy

Group-theoretical symmetry analysis of the tetragonal phase (with space group: I*4/m*) suggests a total of twelve Raman active modes in all these compositions at the zone-center (Γ) with irreducible representation $\Gamma_{Raman}$ = 3 $A_g$ + 3 $B_g$ + 3 $^1E_g$ + 3 $^2E_g$ [17]. Figure 2 (a-c) presents the Raman spectra of all three compositions as a function of temperature which clearly shows the presence of an intriguing broad peak (discussed in the next section), prominent at lower temperatures, along with the narrow phonon bands. As evident from Fig. 2(a), only nine phonon modes (labelled as P1 – P9) could be clearly identified in SCT at 5 K. The remaining modes are not observed likely due to their low Raman scattering cross-section in this Te-based system. On the other hand, Raman spectra of BCW and SCW show additional modes than expected (refer to the spectra 5 K in Fig. 2 (b, c)). As can be seen in Fig. 2, BCW and SCW comprise of a total of 14 (N1 – N14) and 16 (R1 – R16) phonon modes, respectively. A disparity in the observed number of modes with group theory may be attributed to a possible disorder-induced Raman activity of the W-based compositions. A careful analysis of their site-symmetries indicates that the phonon modes observed in these systems originate primarily due to Ba/Sr, O1, and O2 atomic displacements. On the other hand, vibrations due to Cu, W or Te atoms are not allowed to be Raman active as per group-theory. Raman spectra of the three compositions are further analyzed to extract the phonon parameters using the Lorentzian function, the details of which are given in

Table SII in Supplemental Material [16]. The origin of the broad peak observed in our Raman data at low temperatures is discussed below.

### C. Magnetic Raman scattering

Figure 2 displays the Raman spectra for the three systems at a few temperatures showing intriguing phonon features. The Raman spectra collected at 5 K display an unusual broad spectral feature (shaded in orange color) around ~ 270, 196, and 192 cm$^{-1}$ in $Ba_2CuWO_6$, $Sr_2CuWO_6$, and $Sr_2CuTeO_6$, respectively. Such features, in general, may arise due to higher order or multiphonon fast relaxation processes, spin-wave excitations (magnetic scattering), or photoluminescence [16,18-22]. Since the broad feature appears predominantly in the magnetically ordered region (upto $T_S$) for all the compositions, therefore, we attribute such Raman activity to originate from spin-wave excitations. The magnetic scattering-induced Raman activity can stem either from single-magnon (1M) or two-magnon (2M) Raman processes. However, it is easier to distinguish both these magnetic Raman scatterings (1M or 2M) because of a clear difference in their characteristic features and response to the applied magnetic field ($B$). 1M Raman scattering generally originates from the zone-center excitations showing very narrow lineshape and $B$-dependent energy contrary to our observation of broad lineshape [23]. Usually, the origin of scattering from 1M is because of spin-orbit coupling which results in the appearance of a very weak Raman mode, which is not the case here [23]. Therefore, the observed broad modes in all these systems cannot arise due to 1M scattering. Our measurements show that the observed broad features do not exhibit noticeable variation with changing magnetic field and shows a rapid decay with increasing temperature (see Supplemental Material for field dependent Raman data) [16]. Therefore, we attribute the broad feature to arise from scattering due to a pair of magnons (2M) satisfying the conservation law $q + q' = 0$ where $q$ and $q'$ are the magnon wave vectors of an

AFM (see supplemental material for more details) [16, 24]. Investigation of spin-wave excitations in AFMs can render crucial information about the magnetic exchange. For instance, the spectral position ($\omega_{2M}$) of 2M provides an estimate of the exchange interaction ($J$) in Heisenberg antiferromagnets (HAFM) as $\omega_{2M} = J(2zS - 1)$ where $z (= 4)$ refers to the number of nearest magnetic neighbors and $S$ denotes the spin of the magnetic ion [25]. Following this approach, the exchange strength ($J$) is estimated to be ~ 8.16, 8.01, and 11.2 meV for SCT, SCW, and BCW, respectively, consistent with the reported values in earlier studies [9,12]. The estimated $J$'s also corroborate our magnetic data which indicates that magnetic interactions are strongest in BCW. Thus, Raman scattering can act as an alternative probe to confirm the nature of magnetic excitation and estimate the strength of exchange interactions $J$ in quasi-2D antiferromagnets.

Figure 2 presents the temperature evolution of the Raman spectra highlighting the 2M for the three compositions. A comparison among the three compositions suggests that A-site cation significantly influences the spectral position of these excitations, influencing the orbital overlap of the magnetic exchange pathway *via* the B-O-B′ (Cu-O-(Te/W)) bond angle. A careful analysis of bond angles indicates that BCW exhibits a more linear super-exchange pathway (~ 172°) promoting larger orbital overlap and stronger spin interactions than SCW (~ 163°) and SCT (~ 159°) [26]. Therefore, the 2M mode appears at much higher energy (at ~ 269 cm$^{-1}$) in BCW as compared to SCW (~ 215 cm$^{-1}$) and (~ 190 cm$^{-1}$) in SCT. Figure 2 reveals another interesting observation regarding the linewidth of the 2M that largely depends on the nature of the hexavalent (B′) cation. Notably, the $\Gamma$ is large (~ 110 cm$^{-1}$ at 5 K) for W-based double perovskites BCW and SCW than SCT (~ 60 cm$^{-1}$). Since the hexavalent (B′) cations are the exchange mediating agencies along the pathways, they affect the overall lifetime of the excitations by modulating the $J$, eventually influencing the $\Gamma$. The broad linewidth of 2M excitations in these systems is consistent with the conventional characteristic of 2M in $S = \frac{1}{2}$ cuprates [27]. As reported earlier, $S = \frac{1}{2}$ systems exhibit strong intrinsic quantum fluctuations at lower temperatures which predominantly contributes to the broad linewidth of 2M excitations [27]. Strong quantum fluctuations can originate due to magnetic frustration, low dimensionality, or low coordination of the lattice [28], all of which are present in $Ba_2CuWO_6$, $Sr_2CuWO_6$, and $Sr_2CuTeO_6$ to result in such large linewidth of the 2M.

Figure 3 (a-c) presents the evolution of the energy (frequency), linewidth, and intensity of the 2M mode with temperature for all the compositions. Analysis of the 2M parameters reveals a very weak damping as the temperature is raised across $T_N$ up to $T_S$. Importantly, the 2M mode persists even beyond the $T_S$ indicating the low dimensional nature of the spin-spin correlations, as observed in $Sr_2IrO_4$ and $La_2CuO_4$ [29,30]. The rapid decay of this spectral feature upon heating, especially, above the $T_S$, provides further evidence for its identification as spin-wave excitation. The behavior of the 2M mode in BCW reveals an interesting finding *i.e.*, the existence of a broad magnon tail at room temperature and above (~ 400 K) well in the paramagnetic regime, though no such clear magnon tail could be deciphered in SCT and SCW at such high temperatures.

Several descriptions have been provided in the literature to explain magnon tail effects including thermally damped paramagnon or fractionalized excitations, magnetic energy fluctuations, and spin-lattice coupling [31-34]. To be noted that spin-lattice (phonon) coupling becomes evident in a system below the magnetic transition in the spin-ordered regime, thus ruling out its involvement in the origin of broad tail in BCW. Moreover, in general, the tail effect in the Raman spectrum due to magnetic energy fluctuations appears at very low energies near the Rayleigh (excitation) line. We observe the presence of broad magnon tail in BCW at higher temperatures (above the magnetic transition) and also at higher energies (away from the Rayleigh line). Thus, the broad magnon tail in BCW at high temperatures can be assigned to the thermally phonon-damped 'paramagnon excitations'. The evolution of spectral profiles with temperature provides further insights into the observed anomalies in the phonon behavior and an intriguing correlation that develops between the 2M and phonons, as discussed in the next section.

### D. Temperature-dependent Raman studies

Temperature-dependent Raman investigations on three compositions (SCT, SCW, and BCW) present a non-trivial response of phonon frequencies below the magnetic transitions ($T_S$). In general, the renormalization of phonon frequency with temperature due to contributions from interaction with different degrees of freedom can be described as [35,36]

$$\omega(T) = \omega_0 + \Delta\omega_{anh}(T) + \Delta\omega_{el-ph}(T) + \Delta\omega_{sp-ph}(T) \quad (1)$$

where $\omega_0$ is the phonon frequency at absolute 0 K and $\Delta\omega_{anh}(T)$ is the phonon-phonon anharmonic interaction term that carries the contributions from both quasi-harmonicity and intrinsic anharmonicity. $\Delta\omega_{el-ph}(T)$ refers to the electron-phonon interaction depicting the electrical nature of the system and $\Delta\omega_{sp-ph}(T)$ represents the spin-phonon interaction which renormalizes the phonon frequency in the spin-ordered region. Ignoring the higher-order interaction terms in the potential energy, the phenomenological description of the phonon frequency with temperature is given by cubic-anharmonicity ($\omega_{anh}(T)$) which is expressed as [36]

$$\omega_{anh}(T) = \omega_0 + A\left[1 + \frac{2}{\left(e^{\frac{B\omega_0}{T}} - 1\right)}\right] \quad (2)$$

where $\omega_0$ is phonon frequency at 0 K, $A$ is the fitting parameter, and $B = \frac{\hbar}{2k_B}$ ($\hbar$ is reduced Planck constant and $k_B$ is Boltzmann constant). Cubic-anharmonicity is a three-phonon process involving decay of a higher-energy phonon into two lower energy phonons leading to a decrease in frequency with increasing temperature. A departure from this behavior (Eq. 2) occurs due to dominant contributions from phonon-phonon anharmonic ($\Delta\omega_{anh}(T)$), electron-phonon ($\Delta\omega_{el-ph}(T)$) and spin-phonon ($\Delta\omega_{sp-ph}(T)$) interaction terms which is referred as anomalous. In our case, $\Delta\omega_{el-ph}(T)$ is unlikely as all these systems are insulators with large band gap [37,38]. Therefore, the phonon renormalization occurs mainly due to $\Delta\omega_{sp-ph}(T)$ and $\Delta\omega_{anh}(T)$.

### i. Spin-phonon coupling and phonon anomalies

Figure 4 (a-c) shows a few phonon modes that exhibit pronounced anomalies in their thermal response across the short-ranged magnetic transition ($T_S$) in all the compositions. Figure 4 (a) shows that the phonon modes P1, P7, P8, and P9 in SCT exhibit significant departure from the cubic-anharmonic trend below the $T_S$. Similarly, as displayed in Fig. 4(b), the phonon modes R1, R9, R15, and R16 in SCW exhibit a deviation from anharmonic trend below the $T_S$. Notably, similar anomalies have been reported for several other quasi-2D magnets such as $(VO)_2P_2O_7$, iron and ruthenium trihalides, $(CuCl)LaNb_2O_7$ where deviation in phonon response from the anharmonic trend below the magnetic transition temperature is attributed to spin-lattice coupling [39-41]. Therefore, the strong renormalization of the phonon frequency in SCT and SCW near the magnetic transition is an indicative of coupling between spin and phonon degrees of freedom. As opposed to SCT and SCW, BCW shows an intriguing trend in the phonon behavior as depicted in Fig. 4(c). It can be noted that the phonon modes N1 (at ~109 cm$^{-1}$), N2 (at ~116 cm$^{-1}$) and N3 (at ~156 cm$^{-1}$) display normal anharmonic behavior below $T_S$ as the temperature is raised while above $T_S$, N1 shows a weak temperature dependence, whereas, N2 and N3 start to exhibit an anomalous behavior upon heating. Intriguingly, the linewidth of these phonons also exhibits anomalies near the magnetic transitions (see supplemental material) [16]. This renormalization of the phonon self-energy (frequency and linewidth) is again indicative of the presence of spin-phonon coupling. The remaining phonons in these systems nearly follow the expected trend in agreement with the usual anharmonic decay formalism implying weak spin-phonon coupling (see Supplemental Material) [16].

In perovskites, spin-phonon interaction is understood as the atomic displacement modulated super-exchange, mediated *via* the usual B-$O^{2-}$-B or B-$O^{2-}$-B′ (where, B and B′ are usually magnetic cations and $O^{2-}$ is non-magnetic oxygen ion) interaction pathways [42,43]. However, as the B′-site in SCT, SCW, or BCW double perovskites is occupied by Te or W, the exchange pathway between the nearest-neighbor spins gets modified to B-$O^{2-}$-$O^{2-}$-B or B-$O^{2-}$-(Te/W)-$O^{2-}$-B with relatively long-chain interactions that lead to weak spin-phonon coupling. Due to lack of suitable spin models for such long-chain interactions, the strength of spin-phonon coupling (λ) could not be exactly quantified. However, we can understand the λ qualitatively for individual phonons based on its deviation from $\omega_{anh}(T)$ *i.e.,* a larger deviation would imply a larger λ. Figure 4(d) shows that P4 and R4 demonstrate the largest deviation from the anharmonic trend in SCT and SCW, indicating that the strength of spin-phonon coupling (λ) would be the highest for these modes in respective systems.

## ii. Signatures of Magnon-phonon coupling

Another intriguing observation to note with varying temperature is the correlation between the thermal response of the 2M and certain phonon modes (P4 in SCT, R4 in SCW, N2 and N3 in BCW) as depicted in Figs. 4(c) and 4(d). The 2M in SCT shows weak temperature dependence in its frequency upto $T_S$ (~ 80 K) before softening and then disappearing above 120 K (refer Fig. 4(d)). Intriguingly, the P4 phonon of SCT strongly resembles this trend. The response of P4 (at ~ 194 $cm^{-1}$ at 300 K) is noteworthy since it hardly varies in frequency below $T_S$ contrary to the standard expected anharmonic trend with temperature. Above the $T_S$, P4 follows a standard anharmonic trend but with an unusually large redshift in the frequency by 28 cm$^{-1}$ in the range of 5 - 400 K. Since P4 and the 2M have comparable energy and strongly mimic each other in their thermal response, we believe that the two excitations are well-coupled in the magnetically ordered regime (below $T_S$). On the other hand, the phonon mode R4 (at ~ 168 cm$^{-1}$ at 300 K) in SCW exhibits a weak temperature dependence at low temperatures and displays strongly decreasing frequency above the $T_S$ with an unprecedented overall shift of 42 cm$^{-1}$. The 2M behaves in a similar way but softens rapidly with increasing temperature before disappearing above 200 K, indicating a coupling between the 2M and phonon mode R4 in SCW. Thus, the close resemblance between the phonon (P4 and R4) and magnon (2M) response in SCT and SCW can be attributed to magnon-phonon coupling. It is worth to note that both P4 and R4 demonstrate unusually large shift with temperature above their respective $T_S$, indicating a strong anharmonicity which is strongly influenced by the magnon-phonon coupling at low temperatures below $T_S$.

On the other hand, the phonon and magnon behavior of BCW are different from SCT and SCW. It may be noted that the lineshape of the 2M remains mostly symmetric for both SCT and SCW in the entire temperature range before it undergoes a complete decay to yield flattened spectral profile (refer Fig. 2). On the other hand, the lineshape of the 2M in BCW shows intriguing response across the $T_S$ in BCW. It is seen that the 2M remains mostly symmetric with its frequency being nearly temperature-independent below the $T_S$ (~ 110 K) in BCW, above which it develops a significant broadening and shift in frequency. Moreover, the broadened 2M tail of BCW continues to exist even at temperatures above 300 K. Our measurements reveal that the temperature range (above $T_S$) wherein the 2M starts to broaden and redshift significantly, the phonon modes, in particular, N2 (~ 116 cm$^{-1}$) and N3 (~ 156 cm$^{-1}$), commence to exhibit anomalous trend continuing upto 400 K (see Fig. 4(c)). As seen previously, a discrete phonon excitation when overlaps with the energy range of the 2M continuum, it develops a fano asymmetry in the lineshape across the magnetic transition suggesting a coupling between spin and phonon degrees of freedom [44]. However, the scenario in BCW is different since phonons (N2 and N3) exhibit anomalous trend with temperature instead of Fano asymmetry. These phonons exist in close vicinity of the 2M at higher temperatures and therefore, develop a correlation with the 2M. A similar reversal of thermal response in phonon self-energies have been reported in the two-dimensional magnetic system MnPSe$_3$ when the 2M crosses the energy regime of phonon bands, which is attributed to the magnon-phonon hybridization [14]. We believe that a reversal in frequency response of phonons in BCW across the $T_S$ along with linewidth anomalies can, therefore, be ascribed to the magnon-phonon coupling driven by magnetostriction/magnetoelastic effect. Thus, the phonon anomalies in BCW likely originate from the coupling of the direct interaction of lattice and magnon excitations.

### E. DISCUSSION

Pronounced anomalies are evidenced for the phonons which exist close to or below the energy of the magnetic excitation mode (2M). It is to be emphasized here that the phonon modes which couple strongly with the 2M excitation in SCT and SCW show highest spin-phonon coupling in respective systems with strong renormalization in phonon frequency. In BCW, the 2M starts to broaden and redshift rapidly when the phonon modes start to exhibit anomalous trend above $T_S$, indicating a strong correlation between the lattice and spin degrees of freedom. Therefore, we believe that spin-phonon interaction may play a key role in inducing the magnon-phonon correlation *via* magnetostriction/magnetoelastic effects, thus influencing the phonon anharmonicity. Further, these correlations invoke the possible existence of novel hybridized quasi-particles at the intersection of magnon and phonon dispersion in these systems which may be explored in the future by tailoring the lattice/spin degrees of freedom appropriately.

## IV. CONCLUSIONS

To summarize, we explore the magnetic (magnon) and lattice (phonon) excitations and their correlations in square lattice $Ba_2CuWO_6$, $Sr_2CuWO_6$, and $Sr_2CuTeO_6$ antiferromagnets. Two magnetic transitions — short-range ($T_S$) and long-range ($T_N$) spin-orders are revealed from the magnetic measurements for all the compositions. Raman measurements recorded for all the compositions at lowest temperature (~ 5 K) reveals an interesting broad feature insensitive to magnetic field which is assigned to two-magnon (2M) mode. A detailed analysis of thermal response of a few phonon modes (P4, R4, N2 and N3) and the mode 2M in $Sr_2CuWO_6$, $Sr_2CuTeO_6$, and $Ba_2CuWO_6$ clearly indicate the presence of correlation between the magnetic and lattice excitations. Our measurements also reveal a few phonons showing anomalies across the $T_S$ for these compositions implying the presence of spin-phonon coupling. Furthermore, the phonon modes (P4 and R4) which couple strongly with the 2M in $Sr_2CuTeO_6$ and $Sr_2CuWO_6$ exhibit largest spin-phonon coupling in the respective systems with strong renormalization in frequency. All these findings suggest that spin-phonon coupling may be involved in mediating magnon-phonon correlation. These correlations even above $T_S$, thus, indicate towards the plausible existence of paramagnons as well as other potential hybridized quasi-particles which, we believe, will motivate future theoretical and experimental studies.

## V. ACKNOWLEDGMENTS

SS acknowledges Science and Engineering Research Board (SERB) for financial assistance through ECR/2016/001376 and CRG/2019/002668. Support from DST-FIST (Project No. SR/FST/PSI-195/2014(C)) and Nano-mission (Project No. SR/NM/NS-84/2016(C)) are also acknowledged. Authors acknowledge Central Instrumentation Facility at IISER Bhopal for PXRD and SQUID-VSM research facilities. A. J. and A. S. acknowledges the fellowship (09/1020(0179)/2019-EMR-I and 09/1020(0209)/2020-EMR-I) support from CSIR.

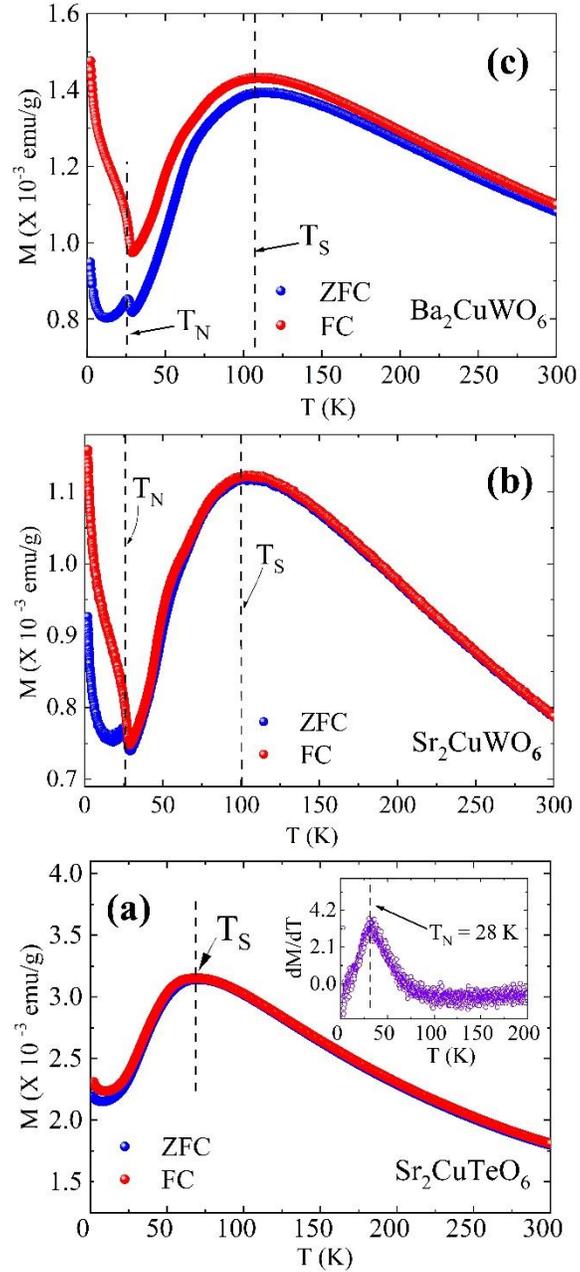

**Figure 1 (Color online).** Temperature-dependent magnetization of (a) $Sr_2CuTeO_6$, (b) $Sr_2CuWO_6$ and (c) $Ba_2CuWO_6$ showing short-range magnetic ordering denoted as $T_S$. $T_N$ represents the long range Néel transition in all the compositions. Inset in the lowermost panel shows derivative of magnetization in $Sr_2CuTeO_6$ to represent the Néel transition.

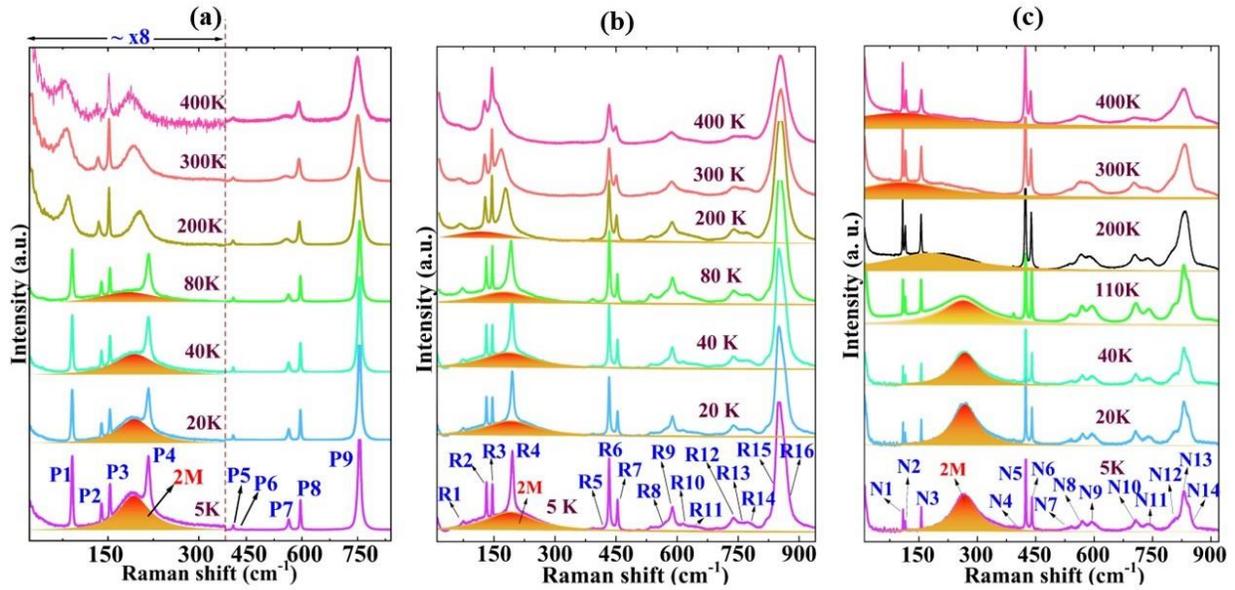

**Figure 2 (Color online).** Evolution of Raman spectra of (a) $Sr_2CuTeO_6$, (b) $Sr_2CuWO_6$ and (c) $Ba_2CuWO_6$ across the magnetic transitions with increasing temperature where phonons are labelled as P1-P9, R1-R16, and N1–N14, respectively. The broad feature shaded with orange color represent the two-magnon (2M) excitation. The spectral range of 10 – 330 $cm^{-1}$ has been magnified for $Sr_2CuTeO_6$ (leftmost panel) for visual clarity.

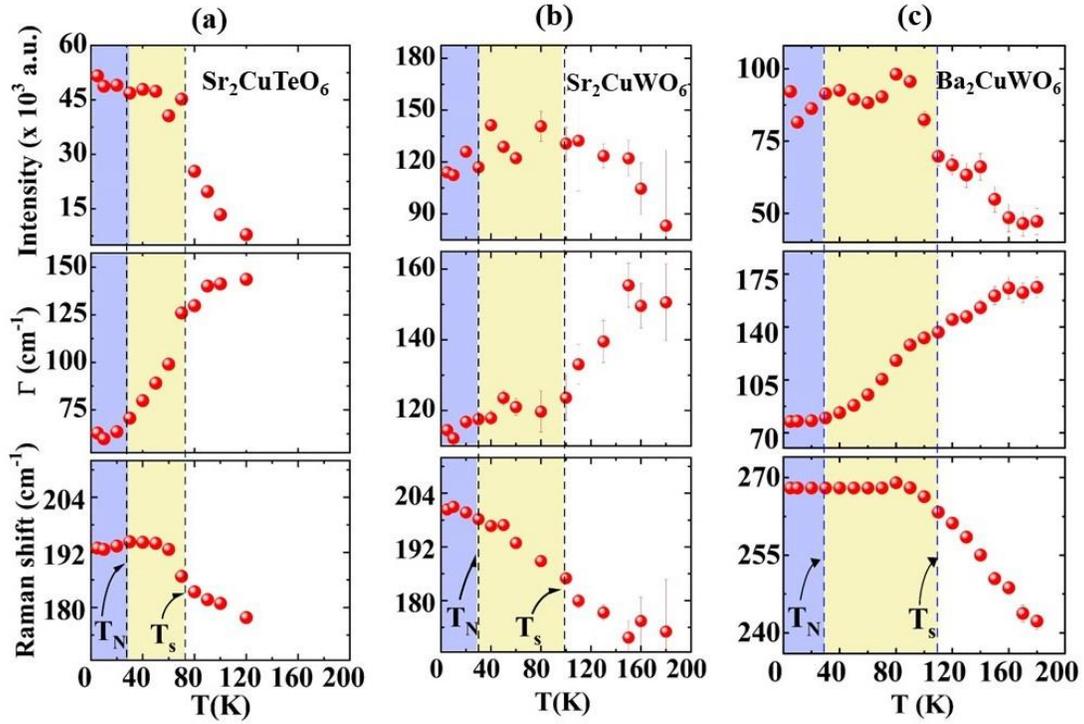

**Figure 3 (Color online)**. Variation of the two-magnon frequency, linewidth ($\Gamma$) and intensity with temperature for (a) $Sr_2CuTeO_6$, (b) $Sr_2CuWO_6$ and (c) $Ba_2CuWO_6$. The shaded region in blue and yellow represents long-range and short-range ordered magnetic regimes, respectively. Error bars are included to show the standard deviation.

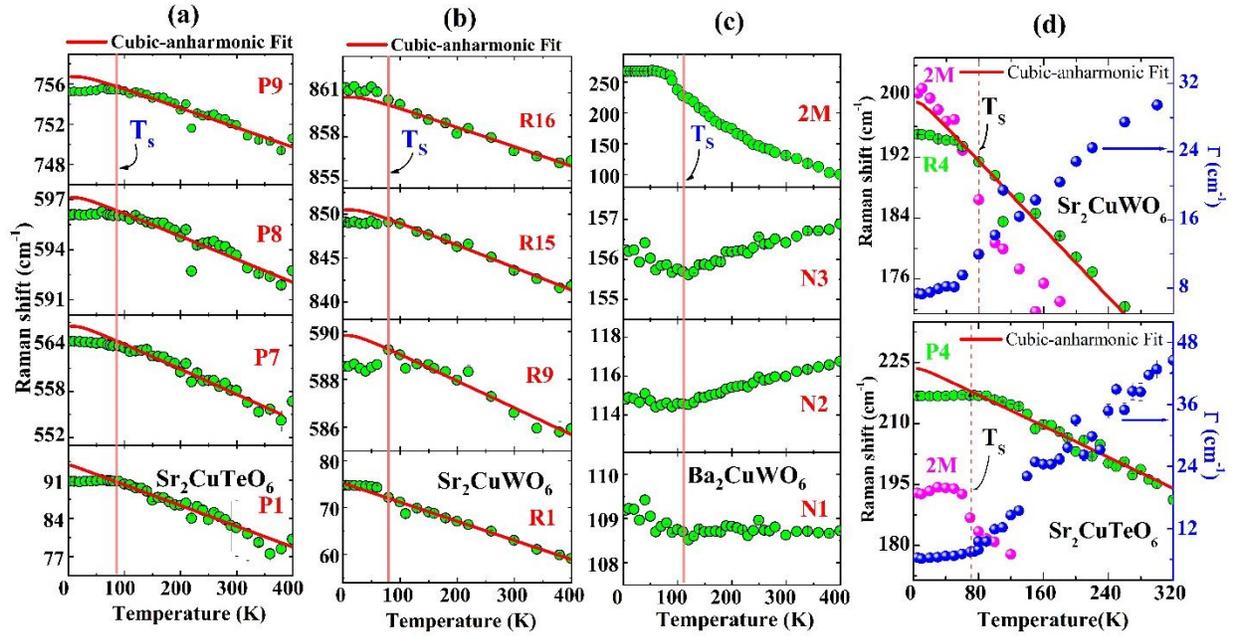

**Figure 4 (Color online)**. Temperature-dependent frequency of the phonons that exhibit deviation from the cubic-anharmonic behavior (shown with solid red line) indicating spin-phonon coupling in (a) $Sr_2CuTeO_6$ and (b) $Sr_2CuWO_6$. (c) Phonons in $Ba_2CuWO_6$ showing opposite thermal response across the $T_S$ along with correlation with the two-magnon (2M). (d) Frequency and linewidth ($\Gamma$) of the phonons P4 and R4 showing coupling with the 2M in respective systems across the magnetic transition ($T_S$). Red vertical lines represent short-range magnetic transition temperature. Error bars are included to represent standard deviation.

# Supplemental Material

# Signatures of novel magnon-phonon coupling in frustrated double perovskite square lattices

Shalini Badola, Aprajita Joshi, Akriti Singh, and Surajit Saha*

*Indian Institute of Science Education and Research Bhopal, Bhopal 462066, India*

*Correspondence: surajit@iiserb.ac.in*

This supplemental material contains room temperature x-ray diffraction data, energy dispersive x-ray data, details of phonon assignments as well as the evolution of Raman spectra with magnetic field and temperature for $A_2CuB'O_6$ (A: Sr, Ba; B': Te, W) *i.e.*, $Sr_2CuTeO_6$, $Sr_2CuWO_6$, and $Ba_2CuWO_6$.

## SI. Room temperature X-Ray diffraction

Figure S1 presents the Rietveld-refined room temperature X-ray diffraction data of $Sr_2CuTeO_6$, $Sr_2CuWO_6$, and $Ba_2CuWO_6$ which confirm that all the systems crystallize into tetragonal symmetry. The obtained lattice parameters are in agreement with earlier studies and are shown in Fig. S1 below [1-3]. Refinement values (goodness of fit *i.e.*, $\chi^2$) are also mentioned for reference.

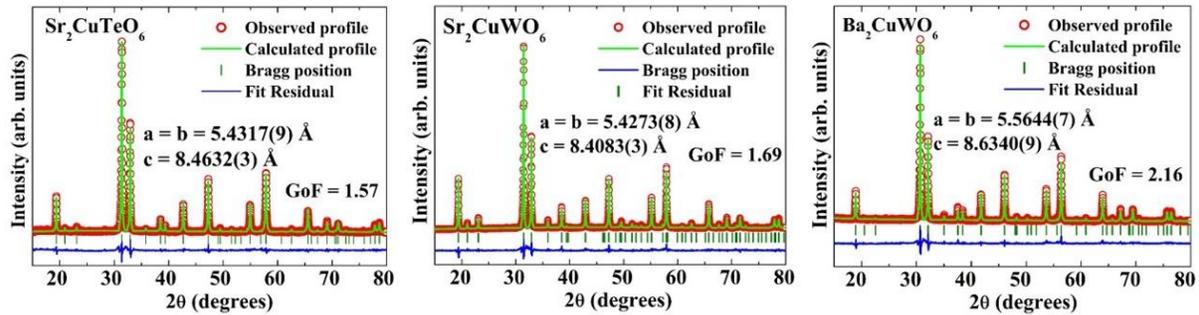

**Figure S1** (Color online): Rietveld refined X-ray diffraction data of $Sr_2CuTeO_6$, $Sr_2CuWO_6$ and $Ba_2CuWO_6$ showing lattice parameters. Open red symbol represents the experimental data, solid green and blue lines are the fitted X-ray profiles and subtracted (the difference of the observed and calculated data), respectively, and the vertical green symbol represent the Bragg lines for $Sr_2CuTeO_6$, $Sr_2CuWO_6$ and $Ba_2CuWO_6$.

## SII. Energy Dispersive X-ray spectroscopy

Energy Dispersive X-ray spectroscopy (EDAX) measurements were carried out on the three double perovskite compositions $Sr_2CuTeO_6$, $Sr_2CuWO_6$, and $Ba_2CuWO_6$ at several spots to confirm the sample stoichiometry. EDAX data reveal that all the elements (particularly the metal ions), are present in the expected proportion within the instrumental limit (~ 2 %) as shown in Fig. S2 (a-c) below. The elemental ratios (Sr/Ba:Cu:Te/W) in $Sr_2CuTeO_6$, $Sr_2CuWO_6$, and $Ba_2CuWO_6$ are 1.91:1.02:1, 1.75:1.25:1, and 2.62:1.39:1, respectively.

**(a)**

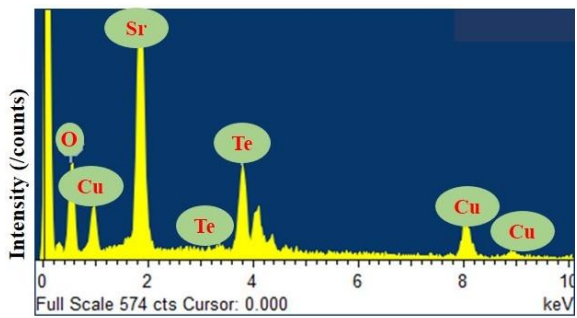

| Spot | Atomic percentage of Elements | | |
|---|---|---|---|
| | Sr | Cu | Te |
| 1 | 16.23 | 8.47 | 8 |
| 2 | 16.24 | 8.60 | 8.42 |
| 3 | 16.41 | 9.26 | 8.86 |
| 4 | 16.46 | 9.11 | 8.88 |
| 5 | 15.49 | 7.64 | 7.98 |
| Average | 16.16 | 8.62 | 8.43 |

**(b)**

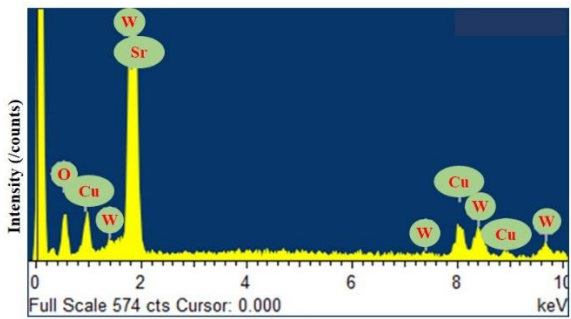

| Spot | Atomic percentage of Elements | | |
|---|---|---|---|
| | Sr | Cu | W |
| 1 | 15.97 | 9.50 | 9.03 |
| 2 | 16.82 | 9.75 | 8.50 |
| 3 | 16.12 | 9.87 | 9.21 |
| 4 | 15.36 | 19.30 | 11.08 |
| 5 | 17.93 | 10.50 | 9.06 |
| Average | 16.44 | 11.78 | 9.38 |

**(c)**

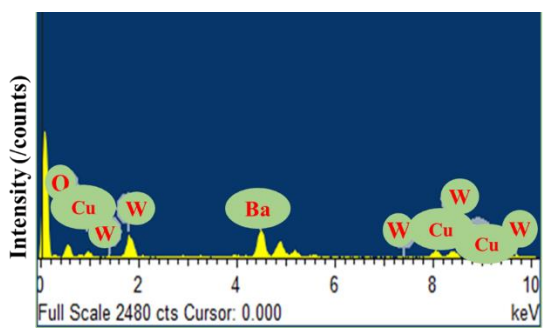

| Spot | Atomic percentage of Elements | | |
|---|---|---|---|
| | Ba | Cu | W |
| 1 | 21.48 | 12.94 | 8.12 |
| 2 | 22.19 | 11.60 | 8.18 |
| 3 | 9.34 | 4.02 | 4.40 |
| 4 | 21.22 | 10.84 | 7.82 |
| 5 | 22.05 | 11.93 | 8.17 |
| Average | 19.25 | 10.26 | 7.34 |

**Figure S2** (Color online): Atomic percentage of elements present in (a) $Sr_2CuTeO_6$, (b) $Sr_2CuWO_6$ and (c) $Ba_2CuWO_6$ along with their Energy Dispersive X-ray spectra.

## SIII. Evolution of spin-orderings in $Sr_2CuTeO_6$, $Sr_2CuWO_6$ and $Ba_2CuWO_6$

Magnetic measurements performed in the temperature range of 2 – 300 K reveal two magnetic transitions; a Néel transition ($T_N$) that appears at lower temperatures while short-range spin ordering ($T_S$) occurs at relatively higher temperatures. While $T_N$ are comparable for all the compositions, $T_S$ differs appreciably. $T_N$ corresponds to a three-dimensional ordering whereas $T_S$ arises due to in-plane (*ab*) spin-spin correlations with highly two-dimensional character. The transition temperatures for all three systems are enlisted in Table SI below. Magnetic data suggest that the spin-spin interactions are weak in $Sr_2CuTeO_6$ as the short-range ordering exhibits narrower maximum in $Sr_2CuTeO_6$. On the other hand, the maxima at $T_S$ are relatively much broader in $Sr_2CuWO_6$ and $Ba_2CuWO_6$ suggesting that the temperature value of 100 K and above is not enough to weaken the in-plane spin-spin correlations significantly in W-based compositions as compared to $Sr_2CuTeO_6$.

**Table SI:** Magnetic transitions of $Sr_2CuTeO_6$, $Sr_2CuWO_6$ and $Ba_2CuWO_6$.

| COMPOSITION | $T_N$ (K) | $T_S$ (K) |
|---|---|---|
| $Sr_2CuTeO_6$ | 28 | 73 |
| $Sr_2CuWO_6$ | 26 | 100 |
| $Ba_2CuWO_6$ | 29 | 110 |

**SIV. Phonons at room temperature**

To investigate the phonon properties, Raman spectra were recorded for all the compositions at room temperature. The phonons are labelled as P1-P9, R1-R16, and N1-N14 for $Sr_2CuTeO_6$, $Sr_2CuWO_6$, and $Ba_2CuWO_6$, respectively, and the corresponding frequencies ($\omega$) are listed in Table SII. The plausible atoms involved in the phonon vibration are also assigned based on the reduced mass ($\mu$) relation with frequency as $\omega \, \alpha \, \sqrt{1/\mu}$.

**Table SII:** Phonon frequencies ($\omega$) of $Sr_2CuTeO_6$, $Sr_2CuWO_6$ and $Ba_2CuWO_6$ at room temperature along with atoms involved in the phonon vibrations.

| $Sr_2CuTeO_6$ | | $Sr_2CuWO_6$ | | $Ba_2CuWO_6$ | | Atoms involved in phonon vibration |
|---|---|---|---|---|---|---|
| Phonon | $\omega$ (cm$^{-1}$) | Phonon | $\omega$ (cm$^{-1}$) | Phonon | $\omega$ (cm$^{-1}$) | |
| P1 | 80.7 ± 0.3 | R1 | 62.8 ± 0.4 | N1 | 108.1 ± 0.1 | Sr/Ba |
| P2 | 133.6 ± 0.4 | R2 | 127.4 ± 0.1 | N2 | 115.6 ± 0.1 | Sr/Ba |
| P3 | 151.7 ± 0.1 | R3 | 145.0 ± 0.1 | N3 | 155.9 ± 0.1 | Sr/Ba |
| P4 | 193.6 ± 0.4 | R4 | 167.8 ± 0.1 | N4 | 393.2 ± 0.1 | O |
| P5 | 414.5 ± 0.2 | R5 | 393.6 ± 2.1 | N5 | 423.2 ± 0.1 | O |
| P6 | 433.0 ± 8.1 | R6 | 433.0 ± 0.1 | N6 | 437.8 ± 0.1 | O |
| P7 | 555.9 ± 0.4 | R7 | 450.6 ± 0.1 | N7 | 532.0 ± 1.1 | O |
| P8 | 592.1 ± 0.1 | R8 | 533.9 ± 1.8 | N8 | 561.9 ± 0.6 | O |
| P9 | 750.6 ± 0.1 | R9 | 586.5 ± 0.3 | N9 | 587.3 ± 1.1 | O |
| | | R10 | 614.5 ± 0.2 | N10 | 702.9 ± 0.4 | O |
| | | R11 | 643.7 ± 5.2 | N11 | 736.3 ± 0.7 | O |
| | | R12 | 739.5 ± 0.4 | N12 | 792.9 ± 0.5 | O |
| | | R13 | 760.2 ± 0.6 | N13 | 826.3 ± 0.5 | O |
| | | R14 | 774.0 ± 0.5 | N14 | 836.5 ± 0.2 | O |
| | | R15 | 843.5 ± 0.2 | | | O |
| | | R16 | 857.0 ± 0.1 | | | O |

**SV. Magnetic field-dependent Raman analysis**

Raman measurements performed at 5 K reveal a broad unusual band (marked in green shade) apart from the narrow phonon lines. Unlike $Ba_2CuWO_6$, this broad feature is found to be missing for

$Sr_2CuTeO_6$ and $Sr_2CuWO_6$ at room temperature (see Fig. 2 in main text). Since this broad feature appears mainly in the magnetically ordered regime, therefore, we attribute it to be of spin origin. In the spin-ordered region, new bands may appear in the Raman spectra mainly due to magnetic scattering from one-magnon (1M) or two-magnon (2M). To confirm whether the feature originates from 1M or 2M scattering, Raman spectra were collected at 5 K at different magnetic fields (0 and 9 T) for all these compositions as shown in Fig. S3. However, no observable change could be deciphered with varying magnetic fields for all the compositions implying that the feature arises due to 2M scattering.

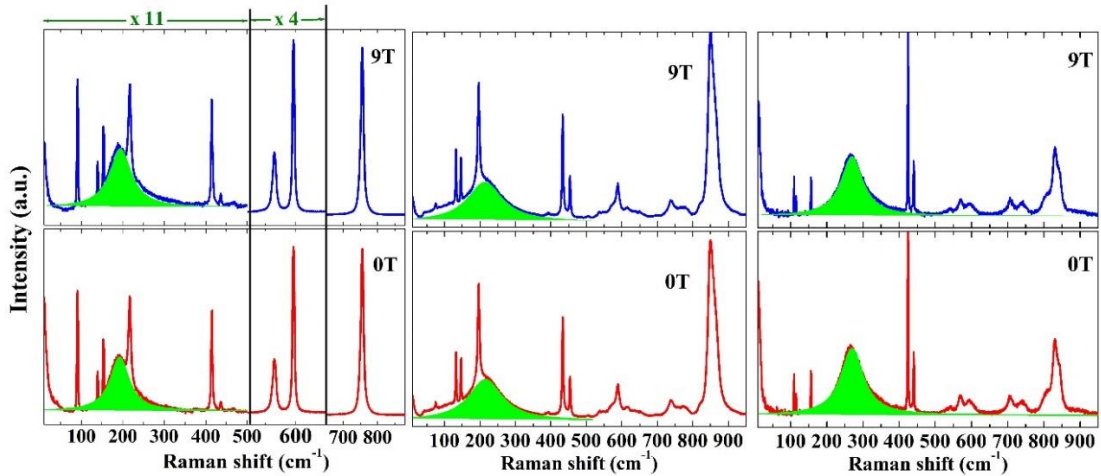

**Figure S3** (Color online): Magnetic field dependent Raman spectra of $Sr_2CuTeO_6$ (left), $Sr_2CuWO_6$ (middle) and $Ba_2CuWO_6$ (right) showing no variation in the two-magnon profile (green shaded profile). The spectral range between 10 - 650 cm$^{-1}$ has been magnified as required in the leftmost panel for visual clarity.

## SVI. Temperature-dependent Raman analysis

Raman analysis reveals a few anomalies in the phonon behavior with temperature. In general, the evolution of phonon frequency with temperature is given by cubic-anharmonicity [4]. A departure from this behavior is observed when phonon interacts with other degrees of freedom such as spin or charge, especially near the transition temperatures. In this study, the renormalization of the phonon mostly occurs across the magnetic transition due to spin-phonon coupling and is shown in the main text in Fig. 4. The remaining phonons obey the standard anharmonic trend and are shown for all the compositions in Figs. S4 and S5 below. To be noted that the Raman measurements were carried out using two different cryostats to span different temperature regimes (5 – 100 K and 80 – 400 K). The measurements performed in two temperature ranges have different optical setups and feedthroughs which lead to a variation in spectral parameters due to different spectral resolutions.

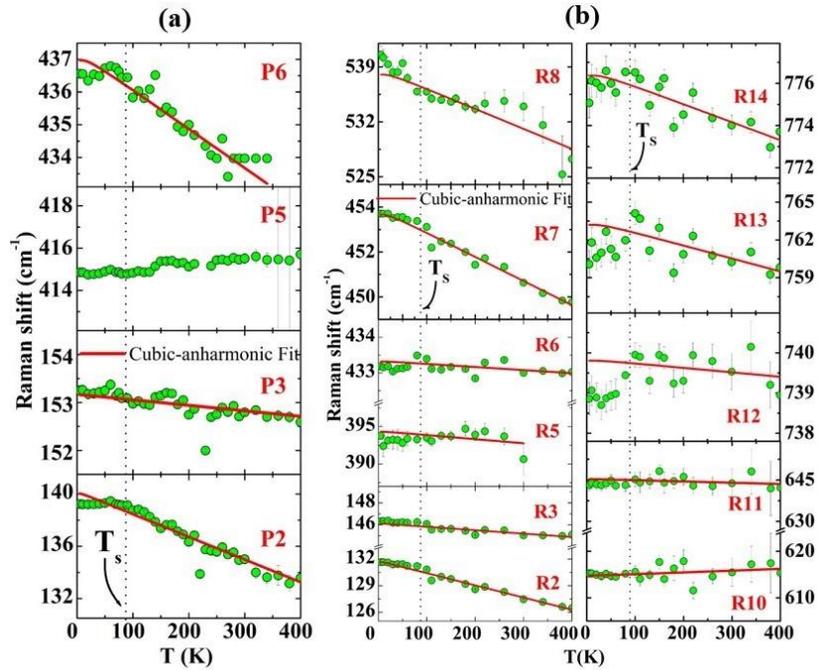

**Figure S4** (Color online): Temperature-dependent phonon frequencies of (a) $Sr_2CuTeO_6$ and (b) $Sr_2CuWO_6$ where most of the phonons follow usual cubic-anharmonic behavior (solid red line). Error bars are included to show standard deviation.

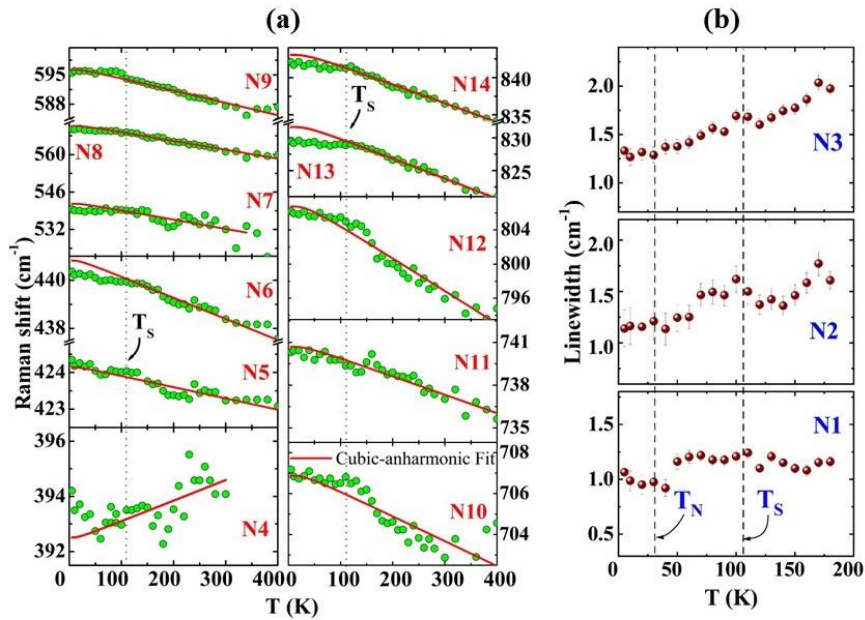

**Figure S5** (Color online): (a) Evolution of phonon frequencies of $Ba_2CuWO_6$ where most of the phonons obey expected Cubic-anharmonic behavior (solid red line). (b) variation of linewidth of phonons in $Ba_2CuWO_6$ which show anomalous thermal response in frequency. Error bars are included to show standard deviation.

**SVII. Raman analysis with variable laser wavelength ($\lambda_{laser}$ = 532nm and 633 nm)**

In order to establish that the origin of broad feature that appear in the magnetically ordered regime of $Sr_2CuTeO_6$, $Sr_2CuWO_6$, and $Ba_2CuWO_6$ is not photoluminescence, we have performed Raman measurement on one of the systems *i.e.*, $Ba_2CuWO_6$ at 80 K (below the short-range ordering temperature $T_S$) with different laser wavelengths ($\lambda_{laser}$ = 532nm and 633 nm). Figure S6 shows the Raman spectra for $Ba_2CuWO_6$ where it can be observed that the broad feature remains invariant in its position (frequency) with different laser wavelengths suggesting that the broad feature does not originate from the luminescence process.

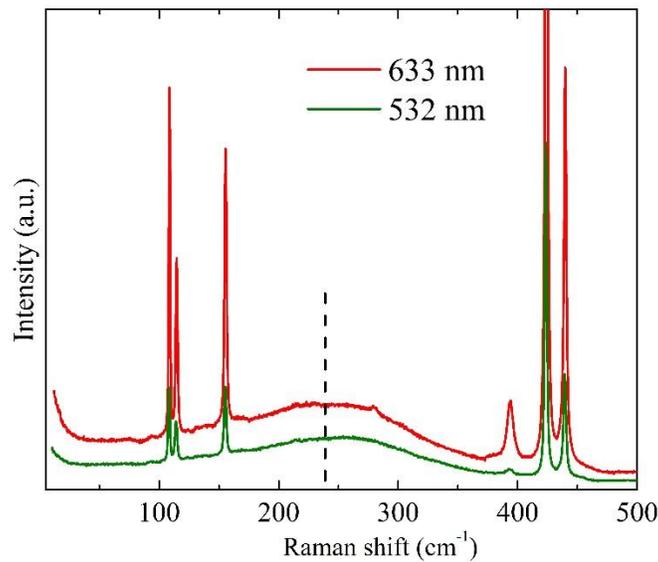

**Figure S6** (Color online): Raman spectra of $Ba_2CuWO_6$ recorded with 532 nm (solid green curve) and 633 nm (solid red curve) laser wavelengths below $T_S$ (at 80 K). Vertical dashed line represents the position of two-magnon.